\documentclass[12pt]{article}
\usepackage{setspace}
\usepackage{amsmath, amssymb}
\usepackage{graphicx}
\usepackage{natbib}
\usepackage{hyperref}
\usepackage{authblk}

\def\N{\mbox{N}}

\newtheorem{prop}{Proposition}
\newtheorem{defn}{Definition}

\usepackage{algorithm}
\usepackage{algorithmicx}
\usepackage{algpseudocode}

\title{Expressing and visualizing model uncertainty in Bayesian variable selection using Cartesian credible sets}
\author{J. E. Griffin\footnote{Department of Statistical Sciences, University College London, Gower Street, London, London WC1E 6BT,
United Kingdom; email: j.griffin@ucl.ac.uk}\\
University College London}
\date{}

\begin{document}

\maketitle

\abstract{
Modern regression applications can involve hundreds or thousands of variables which motivates the use of variable selection methods. Bayesian variable selection defines a posterior distribution on the possible subsets of the variables (which are usually termed models) to express uncertainty about which variables are strongly linked to the response. This can be used to provide Bayesian model averaged predictions or inference, and to understand the relative importance of different variables.  However, there has been little work on meaningful representations of this uncertainty beyond first order summaries. We introduce Cartesian credible sets to address this gap. The elements of these sets are formed by concatenating sub-models defined on each block of a partition of the variables. Investigating these sub-models allow us to understand whether the models in the Cartesian credible set always/never/sometimes include a particular variable or group of variables and provide a useful summary of model uncertainty. We introduce a method to find these sets that emphasizes ease of understanding and can be easily computed from Markov chain Monte Carlo output. The potential of the method is illustrated on  regression problems with both small and large numbers of variables. \\
\\
Keywords: Spike-and-slab prior; Posterior inclusion probability; median model; regression
}

\section{Introduction}

Regression methods allow us to relate the variation in a response to observed variables. If the number of variables are large then 
variable selection methods are needed. Their use is generally motivated by the need to avoid overfitting, which can reduce predictive performance,
 or to understand which subsets of the variables provide parsimonious explanations of the relationship with the response. Many methods have been proposed. Classical approaches include subset selection, stepwise selection methods or penalized maximum likelihood, including Lasso, elastic net, SCAD, or MCP 
\citep[see][for a review]{TaTiWa15}. Bayesian methods place a prior distribution on the regression coefficients leading to a posterior distribution that can express the relative ability of different subsets of the variables to explain the variation in the response. This allows us to understand the relative importance of the variables and the relationship between the variables (for example, that two covariates carry similar information). The approach has been applied widely including linear models \citep{MitBea88, GeoMcC97}, generalized linear models 
\citep{Raf96}
and nonparametric regression \citep{SatVanSha11, LiuRocWan21, LinDu23}.

There are two main classes of Bayesian approach. Firstly, global-local shrinkage priors  encourage coefficients which explain little variation to be shrunk very close to zero
\citep[see {\it e.g.}][for a review]{BhaDatPolWil19}. Post-processing methods can be used to perform variable selection often by finding a sparse summary of the posterior distribution
\citep{HahCar15, PiiPaaVeh20}. Alternatively, spike-and-slab priors place positive probability on some regression coefficients being exactly zero \citep{MitBea88, GeoMcC97}. This leads to a posterior distribution on all possible models defined by subsets of the variables. This approach has good theoretical properties \citep{CasSHVDV15} and predictive performance \citep{PorRaf22}.
Recent work has provided efficient algorithms for use in high-dimensional problems involving tens of thousands of regressors, including Markov chain Monte Carlo (MCMC) methods such as  Importance Tempering \citep{ZanRob19}, ASI 
\citep{GriLatSte21}, LIT \citep{ZhoYan21} 
and
PARNI \citep{LiaLivGri22}, or stochastic search methods such as  SVEN \citep{LiDutRoy23}. We will follow the spike-and-slab approach using MCMC but the methods could be extended to approximations of the posterior from a spike-and-slab prior found using different computational methods such as stochastic search. The inclusion of these priors in a Bayesian hierarchical model allows variable selection on latent variables at different levels of the hierarchy and makes these methods attractive to practitioners.

Bayesian variable selection leads to a posterior distribution on all possible models but inference or prediction usually averages over this posterior distribution to provide summaries or Bayesian model averaged predictions or estimates.     The relative importance of the different variables is often assessed by the posterior probability that a variable is included in the model, which is usually termed the marginal posterior inclusion probability (PIP). The posterior distribution can also be summarized by a single model. The Maximum a posterior (MAP) model is the posterior modal model \citep{RocGeo14}
and the median model \citep{BarBer04}
 includes all variables with a marginal PIP above 0.5. 
 Methods that average over the posterior distribution are feasible even in large spaces using MCMC methods but accurate estimation of posterior model probabilities is usually infeasible if there are many variables.
 
Although, summaries can provide better estimates than a single model \citep[see {\it e.g.}][]{BaBeGeRo21, PorRaf22}, they do not allow us to effectively understand model uncertainty in general.  Credible sets are a promising method for providing more information about model uncertainty in Bayesian variable selection. These guard against overinterpretation of the results of Bayesian variables from considering a single model.
 Initial work has been in the area of Genome Wide Association Studies (GWAS), where Single Nucleotide Polymorphisms (SNPs) are regressed on a trait of interest (such as disease status).  \cite{Mall12} proposed a simple method of finding the credible set for a single regression effect. \cite{WaSaCaSt20} extend this approach to multiple regression effects using the ``sum of single effects'' (SuSiE) approach. This assumes that there are $L$ non-zero regression effects and associates a distribution $\pi_i$ over the SNPs for the $i$-th regression effect, which 
  summarises the uncertainty about the position of the effect. 
   \cite{WaSaCaSt20} provide a method for finding a variational approximation to the posterior distribution of the $\pi_i$'s.
  The method is particularly effective in GWAS where SNPs are serially correlated due to 
  linkage disequilibrium  and it is assumed there is a fixed number of causal SNPs (non-zero regression coefficients).  However, this approach  
 has some drawbacks that can make it less suitable for general regression problems. Firstly, it assumes only negative correlation between the inclusion of variables and so is not suitable for forms of multi-collinearity which lead to positive correlation. Secondly, the value of $L$ must be chosen. Thirdly, credible sets are created for each of the $L$ regression effects rather than the model (which may include different numbers of variables).

In this paper, we take a post-processing approach, which has been successfully used in Bayesian regression \citep{HahCar15, PiiPaaVeh20}, with a spike-and-slab prior
 and work directly on the posterior distribution of all possible models. We  build credible sets on the whole model by approximating  the posterior distribution as a product of blocks of variables and minimizing a criterion that balances the fidelity of the approximation with the complexity of credible sets for each block. This approach allows us to visualize the posterior distribution through the model uncertainty in each block and involves only quantities calculated as averages over models (so making it feasible using MCMC output).

The paper is organized as follows. Section 2 briefly reviews Bayesian variable selection. Section 3 introduces a method for constructing credible model sets with a factorized posterior. Section 4 discusses approximating the posterior distribution with a distribution factorized according to a partition of the variables, and 
choosing the partition to balance the fidelity of the approximation with the complexity of the credible set using an ``ease-of-understanding'' criterion. Section 5 describes visualization of the Cartesian credible sets. Section 6 applies the methods to simulated examples.
Section 7 shows how the method can be  used in different regression problems including a GWAS of Systemic Lupus Erythematosus using a logistic regression model. Section 8 discusses. Appendices include proofs, additional details of the Bayesian methods used in the examples, and further results. Code to implement the method on MCMC samples is available from 
\url{https://jimegriffin.github.io/website/}.

\section{Bayesian variable selection}\label{sec:BVS}

Variable selection methods find a subset of $p$ possible variables which explain the variation in a response through 
a regression model. We refer to variable in the subset as an included variable and 
each possible combination of included variables as a model. Each model can be represented by the $p$-dimensional binary inclusion vector $\gamma$ where the $i$-th variable is only included in the model if $\gamma_i = 1$. Let $\Gamma = \{0, 1\}^p$ be the space of all possible models. Bayesian approaches assume that a prior is placed on $\Gamma$ as well as any parameters of the regression model. 
 \cite{TadVan} provide a book length review  and discuss recent research directions. A main focus of interest is the posterior distribution of the models. In this paper, we do not make any assumptions about the form of regression model or priors used to construct 
 this posterior and 
 we will write it generically as $p(\gamma\mid\mbox{Data})$ for a model $\gamma \in \Gamma$. This is a discrete distribution over $2^p$ possible models and so is hard to 
 visualize or even write down unless $p$ is very small.
 
The large size of $\Gamma$ also motivates the use of computational methods unless $p$ is small. There are two main computational approaches. Firstly, stochastic search methods where $p(\gamma\mid \mbox{Data})$ can be evaluated (for example, linear models) or approximated (for, generalized linear models) for a subset $\Gamma^{\star}$ of $\Gamma$. Several algorithms have been proposed to find a $\Gamma^{\star}$ which provides a good approximation to the posterior distribution
\citep{HanDobWes07,  LiDutRoy23}. The second approach uses MCMC to generate a sample 
$\gamma^{(1)},\dots, \gamma^{(N)}$
 from the posterior distribution $p(\gamma\vert \mbox{Data})$. Unlike the first approach, this leads to unbiased estimates of posterior quantities but it can be computationally expensive to generate a representative sample 
  \citep{ZanRob19, GriLatSte21, ZhoYan21, LiaLivGri22}.

 As discussed in the introduction, 
 there are several methods used to summarize the posterior distribution including:  marginal PIPs, the Maximum a posterior (MAP) model  and the median model \citep{BarBer04, BaBeGeRo21}.
 \cite{BarBer04}  show that the median model is optimal for prediction in orthogonal and nested design (when inclusion variables are 
 independent a posteriori and the PIPs completely characterize the posterior distribution) and can be easily calculated from MCMC output.   If there is correlation between variables then the inclusion variables will be a posteriori correlated and the PIPs do not summarize all information in the posterior. \cite{BaBeGeRo21} discuss properties of the median model under correlated designs. This is a particular problem if the inclusion variables have high correlation which corresponds to multi-collinearity in the design matrix of possible variables. To illustrate this effect,  consider Example 5.2.2 in   \cite{GeoMcC97} where data is simulated from a linear regression model with   $n = 180$ observations and $p = 15$ covariates. The only non-zero 
 coefficients correspond to variables to 1, 3, 5, 7, 8, 11, 12, and 13. There is strong multicollinearity between the following blocks of variables:
 1 and 2; 3 and 4;  5 and 6; 7, 8, 9, and 10; 11, 12 13, 14, and 15. Appendix~\ref{a:GM} provides full details of the data generation.
\begin{figure}[h!]
\begin{center}
\begin{tabular}{cc}
Correlation matrix of variables & Correlation matrix of   $\gamma$\\
\includegraphics[trim = 30mm 90mm 20mm 80mm, scale = 0.4, clip]{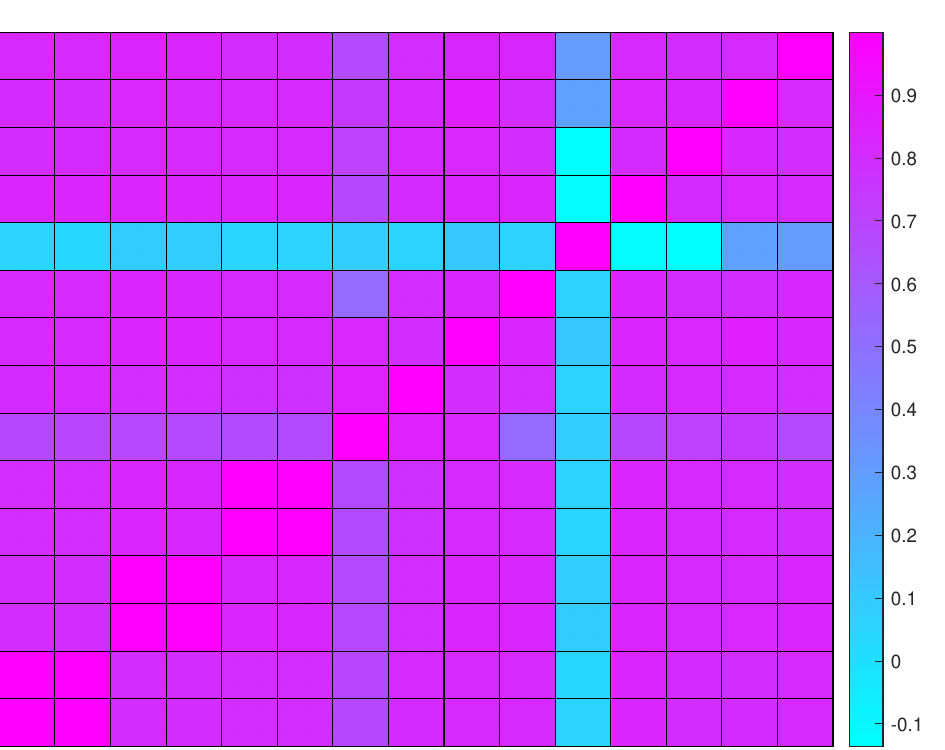} &
\includegraphics[trim = 30mm 90mm 20mm 80mm, scale = 0.4, clip]{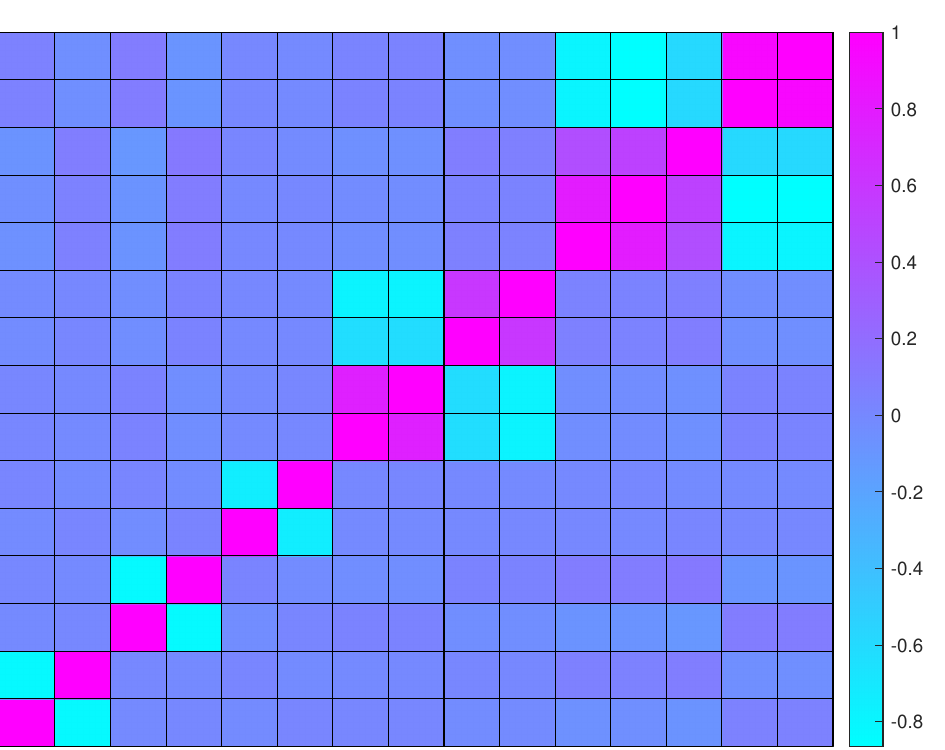}
\end{tabular}
\end{center}
\caption{\cite{GeoMcC97} example: the correlation matrix of the variables and the inclusion variables $\gamma$.}\label{GM1}
\end{figure}
Figure \ref{GM1} shows the correlation matrix of the variables and the posterior correlation matrix of the inclusion variables. The posterior correlation matrix of the inclusion variables shows blocks of correlated variables but with each block uncorrelated. The blocks unsurprisingly mirror the blocks of variables with strong multicollinearity. PIPs and the modal or median model cannot express this correlation information which show strong negative and positive associations. Negative associations are particularly interesting since these show that variables are interchangeable. For example, variables 1 and 2 carry very similar information (similarly for 3 and 4, and 5 and 6).

\section{Cartesian credible sets with factorized posteriors}\label{sec:CS}

\subsection{Credible sets}\label{subsec:CS}

Credible sets  extend credible intervals, which are commonly used to summarize posterior distributions, to posteriors on discrete spaces.
Let $\mathcal{C}$ be the set of all possible subsets of $\Gamma$ then $S\subset \mathcal{C}$ is a $100\,\lambda\%$ credible set if 
$p(S\mid \mbox{Data})\geq \lambda$  \citep{WaSaCaSt20}. The $100\,\lambda\%$ highest posterior density credible interval is the shortest possible 
$100\,\lambda\%$
credible interval and, similarly, we can define the $100\,\lambda\%$ highest posterior probability (HPP) credible set, which is the smallest 
$100\,\lambda\%$ credible set. The $100\,\lambda\%$ HPP credible set can be found by listing the $2^p$ possible models by decreasing probability, 
 \[
p\left(\left.\gamma_{(1)} \right\vert \mbox{Data}\right) \geq 
p\left(\left.\gamma_{(2)} \right\vert \mbox{Data}\right)\geq p\left(\left.\gamma_{(3)} \right\vert \mbox{Data}\right)\geq \dots \geq p\left(\left.\gamma_{(2^p)} \right\vert \mbox{Data}\right)
\]
and
 finding the smallest $K$ such that $\sum_{k=1}^K p\left(\left.\gamma_{(k)} \right\vert \mbox{Data}\right) \geq \lambda$ then 
$\left\{\gamma_{(1)}, \gamma_{(2)}, \dots, \gamma_{(K)}\right\}$ is the HPP credible set.
This approach  is possible if $p$ is relatively small (roughly less than 30) and all posterior probabilities can be calculated or approximated analytically. However, in many cases, $p$ will be larger and can be substantially larger (for example, genomic data sets can have tens of thousands of potential variables) or be part of a larger hierarchical model. In these cases, it's often hard to accurately estimate model probabilities unless MCMC sample sizes are very large, as the level of model uncertainty makes it difficult to visit all plausible models, ({\it i.e.} models with posterior probability greater than $\epsilon$ for $0 < \epsilon < 1$ where, typically, $\epsilon$ is small) using stochastic search or MCMC. Additionally, searching the space of credible sets  $\mathcal{C}$
 is daunting as it contains  
  $2^{2^p}-1$ possible sets and  the size of the HPP credible set can make it hard to write down or consider.

\begin{table}[h!]
\begin{center}
{\footnotesize 
\begin{tabular}{|cc|cc|cc|cccc|ccccc|c|}\hline
1 & 2 & 3 & 4 & 5 & 6 & 7 & 8 & 9 & 10 & 11 & 12 & 13 & 14 & 15 & Prob\\\hline
1 & 0 & 0 & 1 & 1 & 0 & 0 & 0 & 1 &   1 &   0 &   0 &   0 &   1 &   1 & 0.0849\\
1 & 0 & 1 & 0 & 1 & 0 & 0 & 0 & 1 &   1 &   0 &   0 &   0 &   1 &   1 & 0.0817\\
1 & 0 & 0 & 1 & 1 & 0 & 0 & 0 & 1 &   1 &   1 &   1 &   1 &   0 &   0 & 0.0424\\
1 & 0 & 0 & 1 & 1 & 0 & 0 & 0 & 1 &   1 &   0 &   0 &   1 &   1 &   1 & 0.0396\\
0 & 1 & 1 & 0 & 1 & 0 & 0 & 0 & 1 &   1 &   0 &   0 &   0 &   1 &   1 & 0.0345\\
0 & 1 & 0 & 1 & 1 & 0 & 0 & 0 & 1 &   1 &   0 &   0 &   0 &   1 &   1 & 0.0338\\
1 & 0 & 1 & 0 & 1 & 0 & 0 & 0 & 1 &   1 &   0 &   0 &   1 &   1 &   1 & 0.0279\\
0 & 1 & 0 & 1 & 1 & 0 & 0 & 0 & 1 &   1 &   1 &   1 &   1 &   0 &   0 & 0.0264\\
0 & 1 & 0 & 1 & 1 & 0 & 0 & 0 & 1 &   1 &   0 &   0 &   1 &   1 &   1 & 0.0248\\
1 & 0 & 1 & 0 & 1 & 0 & 0 & 0 & 1 &   1 &   1 &   1 &   1 &   0 &   0 & 0.0218\\
1 & 0 & 1 & 0 & 0 & 1 & 0 & 0 & 1 &   1 &   0 &   0 &   0 &   1 &   1 & 0.0218\\
1 & 0 & 0 & 1 & 0 & 1 & 0 & 0 & 1 &   1 &   0 &   0 &   0 &   1 &   1 & 0.0202\\
0 & 1 & 1 & 0 & 1 & 0 & 0 & 0 & 1 &   1 &   0 &   0 &   1 &   1 &   1 & 0.0183\\
1 & 0 & 1 & 1 & 1 & 0 & 0 & 0 & 1 &   1 &   0 &   0 &   0 &   1 &   1 & 0.0176\\
0 & 1 & 1 & 0 & 1 & 0 & 0 & 0 & 1 &   1 &   1 &   1 &   1 &   0 &   0 & 0.0142\\\hline
\end{tabular}}
\end{center}
\caption{\cite{GeoMcC97} example: the models in the 50\% HPP credible set shown with their posterior probability. The vertical lines show the blocks of strongly multicollinear variables.}
\label{GM2}
\end{table}
Before presenting a possible solution, consider the 50\% HPP credible set for Example 5.2.2 of \cite{GeoMcC97}, which is shown in Table~\ref{GM2}. There are 15 models but each model can be constructed by concatenating sub-models $\kappa_1$, $\kappa_2$, $\kappa_3$, $(0, 0, 1, 1)$, and $\kappa_4$ 
to form $(\kappa_1, \kappa_2, 0, 0, 1, 1, \kappa_4)$
where 
$\kappa_1 \in\{(0, 1), (1, 0)\}$, $\kappa_2 \in\{(0, 1), (1, 0)\}$, $\kappa_3 \in\{(0, 1), (1, 0)\}$ and $\kappa_4\in\{(0, 0, 0, 1, 1), (1, 1, 1, 0, 0), (0, 0, 1, 1, 1)\}$. 
This suggests the following interpretation.  Variables 9 and 10  should be included in the model and variables 7 and 8 excluded, only one of variables 1 and 2, one of variables 3 and 4, and one of variables 5 and 6 should be included in the model. Finally, one of the combinations: variables 11, 12 and  13, or variables 13,  14 and 15 or  variables 14 and 15 should be included in the model.  Expressing the models in the credible set as a concatenation of sub-models 
 is easier to understand than enumerating all models in  the 50\% HPP credible set. 
There are 24 models that can be constructed 
in the form  $(\kappa_1, \kappa_2, 0, 0, 1, 1, \kappa_4)$ and this also defines a 50\% credible set. This is larger than the 50\% HPP credible set but much smaller than all possible models ($2^{15} = 32768$). 
 
\subsection{Finding Cartesian credible sets with factorized posteriors}

The  example in subsection~\ref{subsec:CS} suggests that a simpler description of credible sets is possible if we restrict the space of possible credible sets from $\mathcal{C}$ to something more structured. A structure used throughout this paper is the Cartesian product of sets, which we will represent by $\otimes$,
generated using a partition of the variables, which provides a formal way of concatenating elements of different sets. 

\begin{defn}
Suppose that we have $p$ variables and a  partition  $P = \{P_1, \dots, P_L\} $ of $\{1, 2,\dots, p\}$ where $l_i$ is the size of the set $P_i$, 
 $S = \otimes_{i=1}^L S_i\subset \Gamma$ where $S_i$ is a non-empty subset of 
 $\{0, 1\}^{l_i}$ for $i=1,\dots, L$ 
 is a  {\em $100\lambda\%$ Cartesian credible set} if $p(S\mid\mbox{Data}) \geq \lambda$.
\end{defn}

We will refer to each element of the partition as a block,  $l_i$ as the size of the $i$-th block, and $S_i$ as a block credible set.
It is also useful to define $\kappa_i \in \{0, 1\}^{l_i}$ as a sub-model for the $i$-th block. This means that $S_i$ is a set of sub-models $\kappa_i$.
   The choice of a Cartesian product allows the credible set to be understood through the block credible sets in a similar way to our interpretation of the results from the \cite{GeoMcC97} example  above. Imposing this structure leads to $\prod_{i=1}^L (2^{2^{p_i}}-1)$ possible credible sets which is much smaller than all possible credible sets when the maximum of $p_i$ is much smaller than $p$.
The Cartesian product is particularly useful when we assume that the posterior distribution can be factorized over the blocks using the following result.

\begin{prop}
Suppose that $S =  \otimes_{i=1}^L S_i$ and $p(\gamma\mid\mbox{Data}) = \prod_{i=1}^L p(\kappa_i\mid \mbox{Data})$ for all $\kappa_i \in S_i$ for $i = 1,\dots, L$
then
\[
p(S\mid \mbox{Data}) = \prod_{i=1}^L p(S_i\mid \mbox{Data}).
\]
\end{prop}

\subsubsection*{Example 1} Suppose that we have $p=3$ variables, the partition $P = \{1, \{2, 3\}\}$ ($K=2$ blocks) and the posterior distribution in Table~\ref{post1}.
\begin{table}[h!]
\begin{center}
\begin{tabular}{|c|cc|cccc|}\hline
&\multicolumn{2}{|c|}{Block 1} & \multicolumn{4}{c|}{Block 2} \\
&\multicolumn{2}{|c|}{$\gamma_1 = $} & \multicolumn{4}{c|}{$(\gamma_2, \gamma_3) = $} \\
& 0 & 1 &  (0, 0) & (0, 1) & (1, 0) & (1, 1)\\\hline
Posterior probability & 0.9 & 0.1 & 0 & 0.5 & 0.5 &  0\\\hline
\end{tabular}
\end{center}
\caption{Posterior distribution for example 1.}\label{post1}
\end{table}
If we choose $S_1 = \{0, 1\}$ and  $S_2 = \{(0, 0), (0, 1), (1, 0), (1, 1)\}$ then
 \[
 S = \{(0, 0, 0), (0, 0, 1), (0, 1, 0), (0, 1, 1), (1, 0, 0), (1, 0, 1), (1, 1, 0),
 (1, 1, 1)\}. 
\]
The formula follows from noticing that $p(S) = 1$, $p(S_1) = 1$ and $p(S_2) = 1$.
Choosing  $S_1 = \{0\}$ and $S_2 = \{(0, 1), (1, 0)\}$ leads to $S = \{(0, 0, 1), (0, 1, 0)\}$. The formula 
follows from noticing that $p(S) = 0.9$, $p(S_1) = 0.9$, $ p(S_2) = 1$.
\vspace{0.1in}

Conveniently, we try to find small credible intervals but not necessarily the smallest. For example, the commonly used $100\lambda\%$ central credible interval is only the smallest $100\lambda\%$ credible interval if the posterior distribution is symmetric. Similarly, we would like to find a small $100\%\lambda$ Cartesian credible set.  
The smallest $100\,\lambda\%$ Cartesian credible set is the $S_1,\dots, S_L$ which minimize $\sum_{i=1}^L \log\#S_i$ subject to $\sum_{i=1}^L \log P\left(S_i\mid\mbox{Data}\right) \geq \lambda$. By ordering the elements of $S_i$ to have decreasing posterior probability
$p\left(\left.\kappa_i^{(1)}\right\vert \mbox{Data}\right) \leq \dots \leq p\left(\left.\kappa_i^{\left(2^{l_i}\right)}\right\vert \mbox{Data}\right)$
, this can be seen as a nonlinear integer knapsack optimisation problem which is known to be NP-hard in the case of arbitrary non-negative non-decreasing profit functions \citep{GurKopKovKov23}. \cite{GurKopKovKov23} derive approximate algorithms for this problem but these are not invariant to re-ordering of the blocks. This is not suitable here since we wish to treat all blocks in the same way. Therefore, we propose a different heuristic iterative algorithm. Suppose that we have current values 
 $S_1,\dots, S_L$, then removing the sub-model with the smallest marginal posterior probability from each block credible set leads to the smallest reduction in the probability of the Cartesian set, by a factor of  $\frac{ P\left(S_i\mid\mbox{Data}\right) - \min_{\kappa_i \in S_i} P(\kappa_i\mid \mbox{Data})}{\left(S_i\mid\mbox{Data}\right) }$.
 It also leads to the size of the Cartesian credible  set reducing by a factor of $\frac{\# S_i - 1}{\#S_i}$. This suggests using a rule which balances removing a sub-model from the block which has the smallest effect on the probability of the Cartesian credible set with the largest effect on its size.
This is achieved by choosing the block  which leads to the smallest value of the criterion for the $i$-th block
$ \frac{\min_{\kappa_i\in S_i} P(\kappa_i\mid \mbox{Data})/\left(S_i\mid\mbox{Data}\right)}{1/\#S_i}$
and removing the sub-model with the smallest marginal posterior probability.
 This process is terminated when 
$\prod_{i=1}^L P\left(S_i\mid\mbox{Data}\right)  < \lambda$ and the sets for which $\prod_{i=1}^L P\left(S_i\mid\mbox{Data}\right)  \geq \lambda$ are used. 
It is convenient to define  $\Gamma_i = \{0, 1\}^{l_i}$ as the space of all sub-models in the $i$-th block  and to sort the probabilities of the sub-models in ascending order. The algorithm is described in Algorithm~\ref{algo1}.

\begin{algorithm}
\begin{algorithmic}
\State $\pi_i \gets 1$, $r_i = 1$  and $q_i \gets p\left(\left.\kappa^{\left(1\right)}_i \right\vert \mbox{Data}\right)$ 
 for $i= 1, \dots, L$.
\While{$\prod_{i=1}^L \pi_i \geq \lambda$}
\State $k^{\star} \gets \arg\min(q_1, \dots, q_L)$.
\State $\pi_{k^{\star}} \gets \pi_{k^{\star}}  - q_{k^{\star}} $
\State $r_{k^{\star}} = r_{k^{\star}} + 1$, 
\State $q_{k^{\star}} = \frac{p\left(\left.\kappa_{k^{\star}}^{\left(r_{k^{\star}}\right)}\right\vert\mbox{Data}\right)/
\sum_{r = r_{k^{\star}}}^{2^{l_i}} p\left(\left.\kappa_{k^{\star}}^{\left(r\right)}\right\vert\mbox{Data}\right)}{2^{l_i} - r_{k^{\star}}}
$.
\EndWhile

\end{algorithmic}
\caption{Cartesian credible set finding algorithm}\label{algo1}
\end{algorithm}

\section{Cartesian credible sets with general posteriors}\label{sec:FP}

The method developed in Section~\ref{sec:CS} finds a $100\,\lambda\%$ Cartesian credible set if the posterior distribution can be factorized over the blocks in a partition. In many problems, the posterior distribution will not have a factor form and so
 we will choose a factorized approximation  $q(\gamma\vert \theta) = \prod_{i=1}^L q(\kappa_i \mid \theta_i)$ where  $\theta = (\theta_1, \dots, \theta_L)$ are parameters. \cite{PapRos17} consider how a block-diagonal design matrix leads to a factorised form of the posterior distribution and, in contrast to this paper, describe a block-diagonal approximation to the general design matrices.  We first consider how to approximate the posterior distribution for a given partition and then how the partition can be chosen using an ``ease-of-understanding'' criterion which balances fidelity of the approximation with complexity of the credible set. In our approach, we exclude variables which have a marginal PIP below a threshold since these variables will not be included in the Cartesian credible set (we use a threshold of 0.04 in our examples).
 
  The approximation $q$ for a particular partition is chosen by finding $\theta$ which minimize Kullback-Leibler divergence between the posterior and $q$,
\begin{align*}
\mbox{KL} &= \sum p(\gamma\mid \mbox{Data}) \log p(\gamma) - \sum p(\gamma\mid \mbox{Data}) \log q(\gamma) \\
&= C - \sum_{\gamma\in\Gamma} p(\gamma\mid \mbox{Data}) \log q(\gamma\mid \theta) \\
 &= C - \sum_{j=1}^L \sum_{\kappa_j \in \Gamma_j} p(\kappa_j \vert \mbox{Data}) \log q(\kappa_j\mid \theta_j).
\end{align*}
If we can fully evaluate the posterior distribution then this can be calculated analytically.  If there is an MCMC sample $\gamma^{(1)}, \gamma^{(2)}, \dots, \gamma^{(N)} \sim p(\gamma\mid \mbox{Data})$  then we can define $\kappa^{(i)}_j$ to be the values of $\gamma^{(i)}$ restricted to the block $P_j$. If the blocks are relatively small then these sub-model posterior probabilities can be well-approximated using MCMC samples.
A Monte Carlo approximation to the KL divergence can be used
\[
-   \frac{1}{N}\sum_{j=1}^L  \sum_{i=1}^N \log q\left(\left.\kappa_j^{(i)}\right\vert \theta_j \right).
\]
In this paper, we define probabilities $q(\kappa_i = \phi)  = \theta_i(\phi)$ for $\phi \in \Gamma_i$. The optimal choice under KL divergence is $\theta_j(\phi) = \frac{1}{N} \#\left\{\kappa_j^{(i)} = \phi\right\}$. Alternatively, we could use a parametric distribution such the 
quadratic exponential binary model \citep{CoxWei94} or the more general
 multivariate Bernoulli distribution \citep{Dai13}.

We use an agglomerative algorithm based on the KL divergence between the posterior and the factorized approximation to choose the tuning parameters which define the partition.
Let $\mbox{KL}(P)$ be a partition $P$ to be the Kullback-Leibler divergence between the posterior distribution and the approximation $q$ calculated using the optimal values of the parameters for $P$. A sequence of partitions can be calculated by merging two blocks of the partition at the previous iteration (in a similar way to agglomerative clustering). 
The two blocks to be merged are chosen to be ones that lead to the largest change in $\mbox{KL}(P)$. We define $\eta_i$ to be the change in $\mbox{KL}(P)$ at the $i$-th iteration of the algorithm.
This greedy approach will lead to the largest improvement in the approximation at each stage
and so defines a sequence of partitions $P_1,\dots, P_L$ which are increasingly close approximation of the posterior distribution indexed by $\eta_1, \dots, \eta_L$.
 The algorithm is summarized in Algorithm~\ref{algo2}.

\begin{algorithm}
 \begin{algorithmic}
\State $P_0 = \{\{1\}, \{2\}, \dots, \{p\}\}$.
\For{$k = 1,\dots, p$}
\State Calculate $D_{i, j} 
 = \mbox{KL}(P_{i, j})$ where $P_{i, j}$ is the partition joining the $i$-th and $j$-th element of $P$.
\State Find $(i, j)$ which maximizes $D_{i, j}$ and let $a_k$ be this value.
\State Set $P_k$ as $P_{k-1}$ with elements $i$ and $j$ joined and 
\State Set $\eta_k = KL(P_k) - KL(P_{k-1})$.
\State Set $S^{(k)}$ as  the $100\lambda\%$ Cartesian credible set found using Algorithm~\ref{algo1}
\EndFor

\end{algorithmic}
\caption{Agglomerative algorithm}\label{algo2}
\end{algorithm}

We could choose the tuning parameters using a criterion which measures the discrepancy between the approximation and the posterior (for example, using Kullback-Leibler divergence). However, this may lead to Cartesian credible sets with large block credible sets. Therefore, we use an alternative criterion which balances  the fidelity of the approximation to the posterior distribution and the  ``ease-of-understanding'' of the credible set. It is hard, or perhaps impossible, to provide a precise definition of  ``ease-of-understanding'' of the credible set but we take the view that smaller block credible sets makes interpretation simpler and so the largest block credible set shouldn't be ``too large''. 
To achieve this, we  use an ``ease-of-understanding'' penalized minimization criterion to the sequence of Cartesian credible sets define by $\eta_1, \dots, \eta_J$.  Our criterion is
\[
\log \#S^{(j)} - f(S^{(j)})
\]
where $f$ is a function of the credible set $S^{(j)}$ and penalizes partition leading to harder to understand credible sets. The first term measure the fidelity of the 
 approximation since a partition with a smaller number of larger blocks will lead to better fidelity of the approximation and a smaller credible set. We choose to define the penalty 
on the partitions $P^{(j)}$ rather than the credible sets $S^{(j)}$. This is still challenging. We use a form  inspired by the ``rich-get-richer'' property of the Dirichlet process \citep{Ferg73} and penalizes partitions with large blocks. The penalty is  the logarithm of the exchangeable product partition function of the Dirichlet process 
\[
f(S^{(j)})
= K^{(j)}\log M +  \sum_{k=1}^{K^{(j)}} \log \Gamma\left(\#P_i^{(j)}\right).
\]
The parameter $M$ controls the level of penalization and we have found that $M=2$ works well in practice.

\section{Interpreting and visualizing  the $100\lambda\%$ Cartesian credible set}

The $100\lambda\%$ Cartesian credible sets and the approximation of the posterior distribution as a product of distribution over blocks of variables
give us a convenient way to interpret the model uncertainty expressed by the posterior distribution of $\gamma$. The product structure implies that the inclusion of blocks of variables are independent and so each block can be interpreted as a separate effect in the regression model. We define the {\it block PIP} for the $i$-th block to be $p(S_i - {\bf 0}_i\mid \mbox{Data})$
(where $-$ represents set difference
and ${\bf 0}_i = \underbrace{(0, \dots, 0)}_{l_i\mbox{  times}}$), which is the posterior probability that at least one variable in the block is included in the model. 
We also define the {\it modal sub-model} for the $i$-th block to be  the sub-model with highest marginal posterior probability in  block $i$.
It is natural to consider the importance of a block of variables by considering whether the block PIP is larger than a threshold (for example, 0.5). If all blocks are singletons, the $\gamma_i$'a are independent under the approximation and this leads to the median model if the threshold is chosen to be 0.5. After deciding which blocks contain important variables, we  consider the marginal posterior distribution of the sub-models in the important blocks.  If the block is a singleton, the block PIP is just the marginal PIP and so that variable can be considered important. If the block is not a singleton,  the posterior distribution of sub-models in a block can range from being very concentrated on the modal  sub-model to being spread over a range of sub-models. Correspondingly, our interpretation can range from considering only the modal sub-model to being unable to choose between all or several sub-models.

 To illustrate this method of interpretation and how the sets can be visualised, we find the $50\%$ Cartesian credible set with simulated data generated using the set-up in Example 5.2.2 of \cite{GeoMcC97},  Recall that there 
are $p = 15$ covariates. The only non-zero 
 coefficients correspond to variables to 1, 3, 5, 7, 8, 11, 12, and 13. There is strong multicollinearity between the following blocks of variables:
 1 and 2; 3 and 4;  5 and 6; 7, 8, 9, and 10; 11, 12 13, 14, and 15.  Appendix~\ref{appendix:setup} gives prior settings and the MCMC scheme.
 \begin{figure}[h!]
\begin{center}
\begin{tabular}{c}
PIPs \\
\includegraphics[scale = 0.8, trim = 25mm 135mm 30mm 135mm, scale = 1, clip]{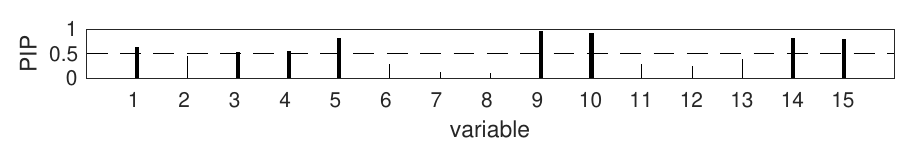}\\
50\% Cartesian credible set\\
\hspace{0.5in} \includegraphics[scale = 0.8, trim = 35mm 140mm 30mm 135mm, scale = 1, clip]{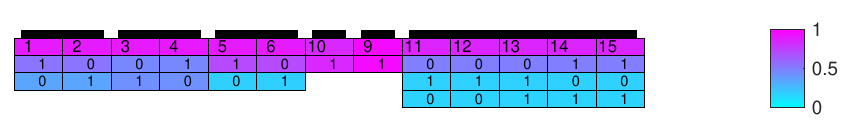}
 \end{tabular}
\end{center}
\caption{\cite{GeoMcC97} example: Top row: PIPs for each variable with a thicker line indicating variables included in the median model. Bottom row:
 $50\%$ Cartesian credible set with $M=2$. Black lines indicate the blocks of the partition.}\label{fig:GM_final}
\end{figure}
A graphical representation of the $50\%$ Cartesian credible set is shown in the bottom row of Figure~\ref{fig:GM_final}. In this graph,
the top row shows the variables which are included in at least one sub-model in the set. Here, variables 7 and 8 are excluded. Black lines above the variable names show the block. In this case, the block are $\{1, 2\}$, $\{3, 4\}$, $\{5, 6\}$, $\{10\}$, $\{9\}$, and $\{11, 12, 13, 14, 15\}$. These are similar to grouping of variables with strong  multicollinearity
identified  in Section~\ref{sec:BVS} with the exception that 
 $\{7, 8, 9, 10\}$ is split into two blocks $\{10\}$ and $\{9\}$ and $7$ and $8$ are completely removed. The boxes around the variable numbers are coloured by the block PIP. All blocks have high block PIPs. Therefore, we can conclude that at least one of the variables in each block is important in the regression.
 
 The sub-models in each block credible set are listed below the variables in the block. 
For example the sub-models for the first block are  $S_1 = \{(0, 1), (1, 0)\}$.
This suggests that either 1 or 2 are included (but not both). Each sub-model is  coloured according to their marginal posterior probability ($p(\kappa_i \mid \mbox{Data})$   for a sub-model $\kappa_i$).  These are 0.38 for  $(0, 1)$ and 0.52 for $(1, 0)$. This can be interpreted as evidence that one of  1 or  2 should be included in the model but there is not strong evidence for preferring one variable over the other (although, the posterior probability slightly favours variable 1). We can draw similar conclusions about the pairs of variables:  3 and 4, and  5 and 6. For
the block $\{11, 12, 13, 14, 15\}$, there are three plausible submodels. One of these is the true subset of important variables which includes 11, 12 and 13. However, the marginal posterior probability of this submodel is only 0.18 and the two other models have stronger or similar support (0.51 and 0.17). 
 The 50\% Cartesian credible set has  24 models compared to 15 models in the 50\% HPP credible set and the 50\% HPP credible set is a subset of the 50\% Cartesian credible set. The difference is small compared with the total number of models (32768).

It's useful to compare the 50\% Cartesian credible set to the median model which includes 1, 3, 4, 5, 9, 10, 14 and 15. This model includes the correct number of variables (8) but includes 5 variables which are not actually important (4, 9, 10, 14, 15). The 50\% Cartesian credible set provides a more nuanced summary. Firstly, the median model chooses both 3 and 4 when the posterior distribution shows clear support for one (but not both) being included. Secondly, the median model excludes the truly important variables 11, 12 and 13 and selects  the (correlated) variables 14 and 15. The 50\% Cartesian credible set shows that at least one of $\{11, 12 , 13, 14, 15\}$ should be included and there are three plausible sub-models. One of these sub-models is the true combination of variables $(1, 1, 1, 0, 0)$. The median model  is forced to choose a single model and ignores information about the relationship between variables.  The Cartesian credible set approach allows us to visualize the relationship between variables and to concentrate on collections of plausible sub-models. This avoids the dangers of overinterpreting the median model (particularly for variables with PIPs close to 0.5) and can be used to find, for example,  the interchangeability of variables and allows us to address 
dilution effects \citep{Geo10}, the splitting of posterior probability between models including subsets of a set of strongly correlated variables.

\section{Simulated example: the effects of sample size in block correlated example}

To illustrate how the Cartesian credible set changes with sample size, we consider the following simulated example. Data are simulated from a linear regression with $p=15$ regressors and $n=640$ observations where
\[
y_i = X_i \beta + \epsilon_i
\]
with $\epsilon_i \sim \N(0, \sigma^2)$, $X_i \sim \N(0, A(\rho))$ and $\beta = (1, -1, 0, 1, 0, 0, -1, 0, 0, 0, 1, 0, 0, 0, 0)$ (so that the 1st, 2nd, 4th, 7th and 11th regression coefficients are non-zero). 
The $p\times p$ correlation matrix of the regressors is block diagonal leading to correlation between regressors in the same block and has the form
\[
A(\rho) = 
\left(
\begin{array}{ccccc}
A_1(\rho) & 0 & \hdots & 0\\
0 & A_2(\rho) & \hdots & 0 \\
\vdots & \vdots & \ddots & \\
0 & 0 & & A_5(\rho)
\end{array}
\right)
\] 
where the $(j, k)$-th element in $A_i(\rho)$ is $\rho^{\vert j - k\vert}$and  higher values of $\rho$ lead to stronger correlations between regressors in the same block. The choice of regression coefficients leads to one (and only one) non-zero regression coefficient in each block. We would expect that it is harder to distinguish the effects of variables in the same block as $\rho$ increases
 and that this effect decreases as $n$ increases. The prior settings and MCMC scheme used for inference are given in Appendix~\ref{appendix:setup}.

We fit a linear regression model to the data using the first $n$ observations for $n=40$, $n=80$, $n=160$, $n=320$ and $n=640$.
\begin{figure}[h!]
\begin{tabular}{ccc}
$n$ & PIP & CCS\\
 \raisebox{0.4in}{40} & 
\includegraphics[scale = 0.5, trim = 45mm 130mm 45mm 130mm, clip]{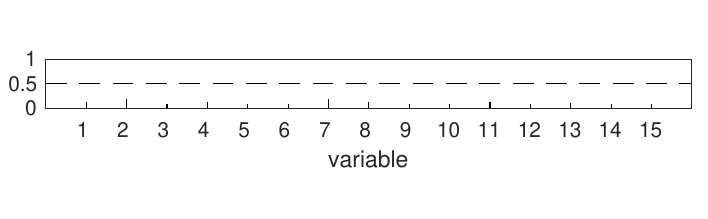} &
\includegraphics[scale = 0.5, trim = 30mm 130mm 30mm 130mm, clip]{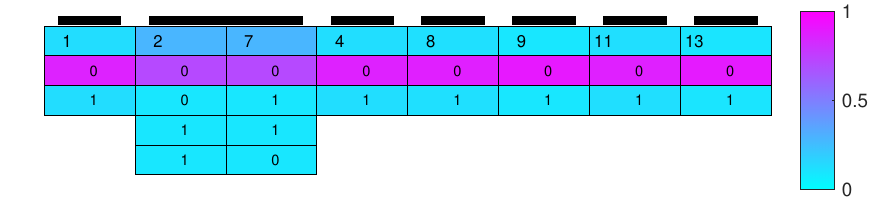}\\
\raisebox{0.4in}{80} &
\includegraphics[scale = 0.5, trim = 45mm 130mm 45mm 130mm, clip]{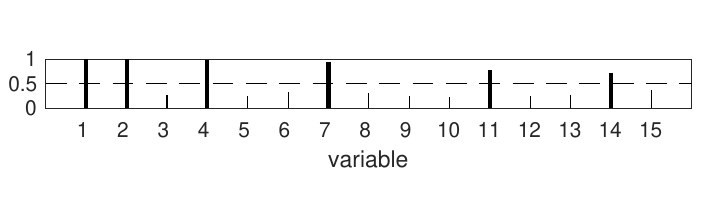} &
\includegraphics[scale = 0.5, trim = 30mm 130mm 30mm 130mm, clip]{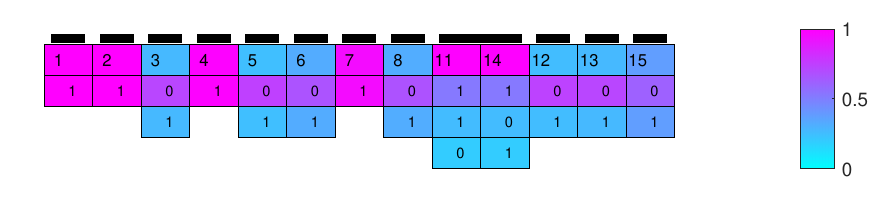}\\
\raisebox{0.4in}{160} &
\includegraphics[scale = 0.5, trim = 45mm 130mm 45mm 130mm, clip]{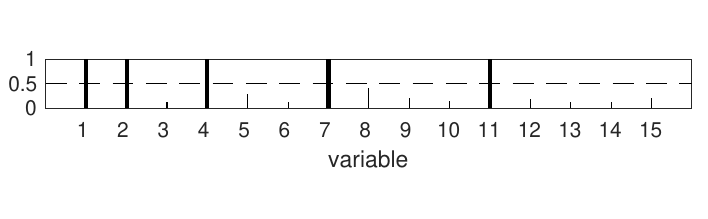} &
\includegraphics[scale = 0.5, trim = 30mm 130mm 30mm 130mm, clip]{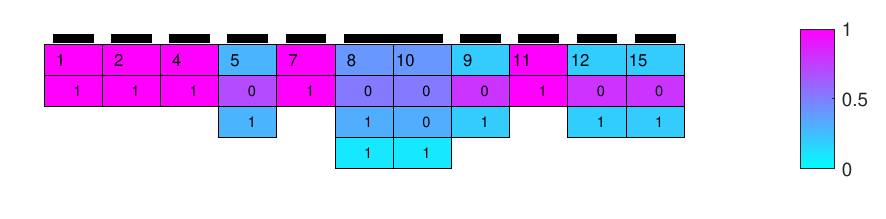}\\
\raisebox{0.4in}{320} &
\includegraphics[scale = 0.5, trim = 45mm 130mm 45mm 130mm, clip]{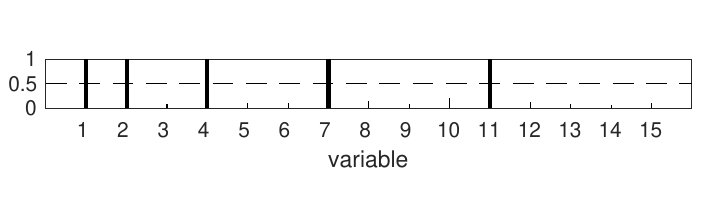} &
\includegraphics[scale = 0.5, trim = 30mm 130mm 30mm 130mm, clip]{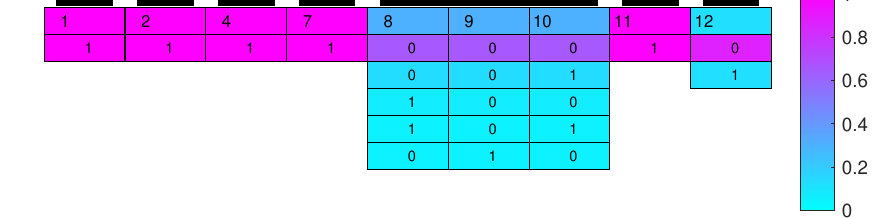}\\
\raisebox{0.4in}{640} &
\includegraphics[scale = 0.5, trim = 45mm 130mm 45mm 130mm, clip]{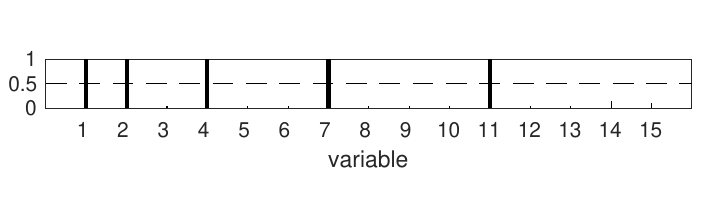} &
\includegraphics[scale = 0.5, trim = 30mm 130mm 30mm 130mm, clip]{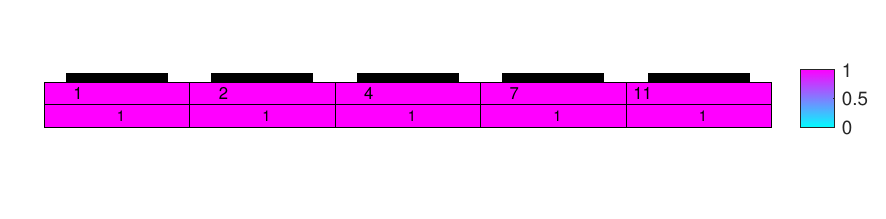}
\end{tabular}
\caption{Simulated example:  PIPs and $50\%$ Cartesian credible sets (CCS) with $M=2$ for simulated datasets with $\rho = 0.5$  and different sample sizes. The variables in the median model are indicated by thicker lines in the PIP plot.}\label{f:sim21}
\end{figure}
The 50\% Cartesian credible sets are presented in Figure~\ref{f:sim21} for $\rho=0.5$. As expected, the important variables are mostly clear for a sufficiently large value of $n$ ($n=80$ in this case). When $n=80$, 1, 2, 4, and 7 are included as singleton blocks which each have high PIPs. The other important variable, 11, is included in a block with regression coefficient 14 which expresses uncertainty about which variable is important. The median model includes both 11 and 14 
 but the graphical representation of the 50\% Cartesian credible set shows that this is because the posterior probability that both variables are included is 0.53 and there is also support for only one of the variables being included.
 For $n=160$ and $n=320$, blocks of unimportant variables are identified but in total these blocks have block PIPs  below 0.5 which suggests that 
 these groups of variables are not important.

\begin{figure}[h!]
\begin{tabular}{ccc}
$n$ & PIP & CCS\\
 \raisebox{0.4in}{40} & 
\includegraphics[scale = 0.5, trim = 45mm 130mm 45mm 130mm, clip]{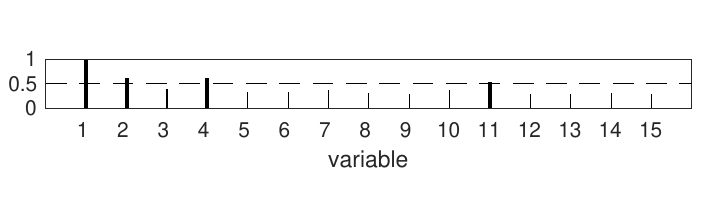} &
\includegraphics[scale = 0.5, trim = 30mm 130mm 30mm 130mm, clip]{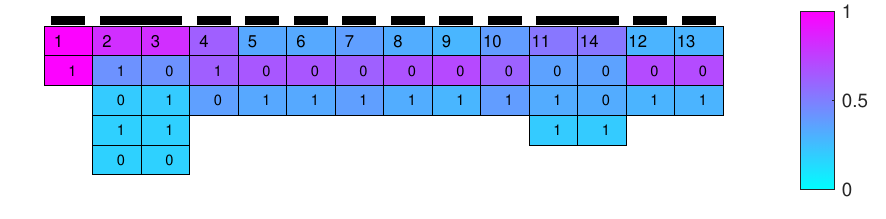}\\
\raisebox{0.4in}{80} &
\includegraphics[scale = 0.5, trim = 45mm 130mm 45mm 130mm, clip]{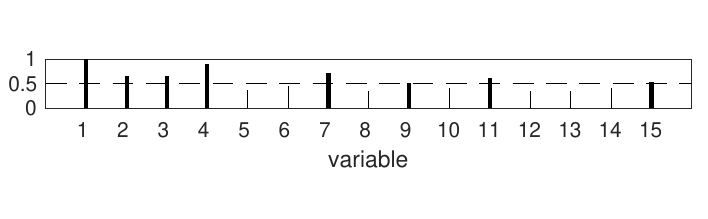} &
\includegraphics[scale = 0.5, trim = 30mm 130mm 30mm 130mm, clip]{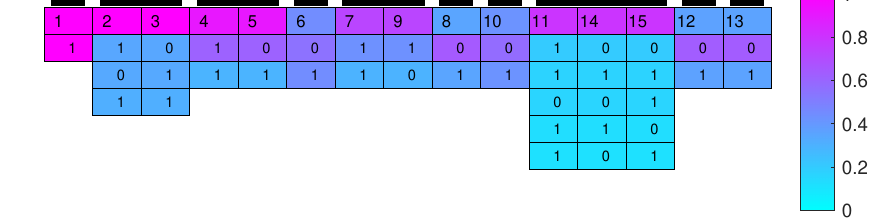}\\
\raisebox{0.4in}{160} &
\includegraphics[scale = 0.5, trim = 45mm 130mm 45mm 130mm, clip]{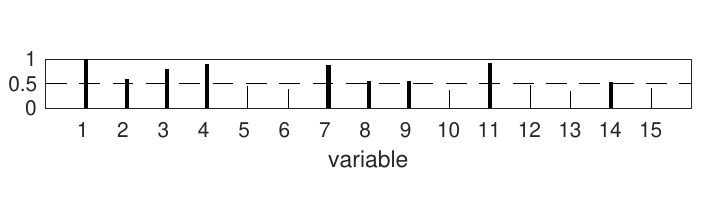} &
\includegraphics[scale = 0.5, trim = 30mm 130mm 30mm 130mm, clip]{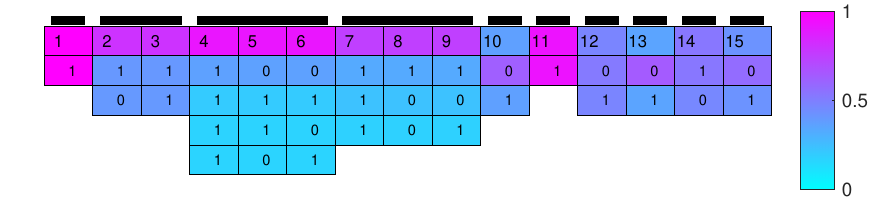}\\
\raisebox{0.4in}{320} &
\includegraphics[scale = 0.5, trim = 45mm 130mm 45mm 130mm, clip]{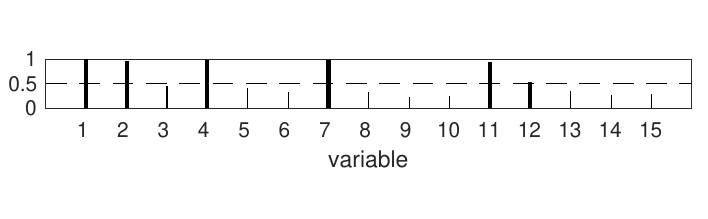} &
\includegraphics[scale = 0.5, trim = 30mm 130mm 30mm 130mm, clip]{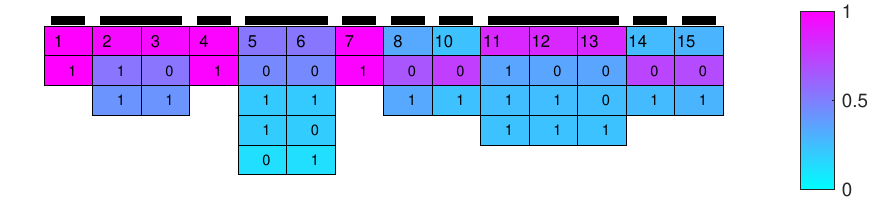}\\
\raisebox{0.4in}{640} &
\includegraphics[scale = 0.5, trim = 45mm 130mm 45mm 130mm, clip]{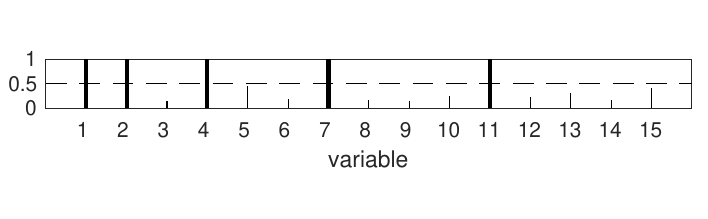} &
\includegraphics[scale = 0.5, trim = 30mm 130mm 30mm 130mm, clip]{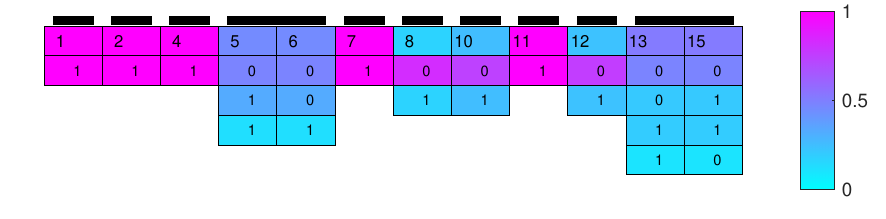}
\end{tabular}
\caption{Example 2: PIPs and $50\%$  Cartesian credible sets (CCS) with $M=2$
for simulated datasets with $\rho = 0.9$  and different sample sizes. The variables in the median model are indicated by thicker lines in the PIP plot.}\label{f:sim22}
\end{figure}
The results with $\rho = 0.9$ are shown in Figure~\ref{f:sim22}. As expected, the larger level of correlation within blocks of variables makes the variable selection problem harder. The median model tends to choose too many variables. For example, 3, 9, and 15 are included in the median model when $n=80$ and 3, 8, 9, and 14 when $n=160$. The underselection disappears as $n$ increases and the five truly important variables are selected when $n=640$. This underselection is due to the correlation in the regressors leading to more than one variable in a block being selected.
The 50\% Cartesian credible set allows us to understand this. For example, when $n=80$, (2, 3), (7, 9) and (11, 14, 15) are indicated in blocks. This shows a truer picture of the data generating mechanism that one of 2 or 3, one of 7 and 9, and one 11, 14 and 15 should be included in the model. We see a similar effect when $n=160$, the blocks again show the relationships between variables 2 and 3, and 7, 8, 9. However, variable 14 is not included in a block with variable 11 since variable 11 has a high PIP. When $n=320$ and $n=640$, the posterior concentrates posterior probability on the true model (and very similar models) and so the ambiguity about the important variables disappears as again illustrated by the Cartesian credible sets.

\section{Applications}

We consider four applications to illustrate the use of Cartesian credible sets in real problems. The first three examples use linear regression models to analyse data  from the \href{https://archive.ics.uci.edu}{UCI Machine Learning Repository} and the fourth example uses a logistic regression model to analyse data from a genome-wide association study (GWAS) of Systemic Lupus Erythematosus. The prior settings and MCMC schemes used for inference are given in Appendix~\ref{appendix:setup}.

\subsection{Behaviour of traffic in Sao Paulo, Brazil}

The data describe traffic levels in Sao Paulo, Brazil from 14th to 18th December 2009. 
\begin{figure}[h!]
\begin{center}
\includegraphics[scale = 0.8, trim = 30mm 85mm 30mm 80mm, clip]{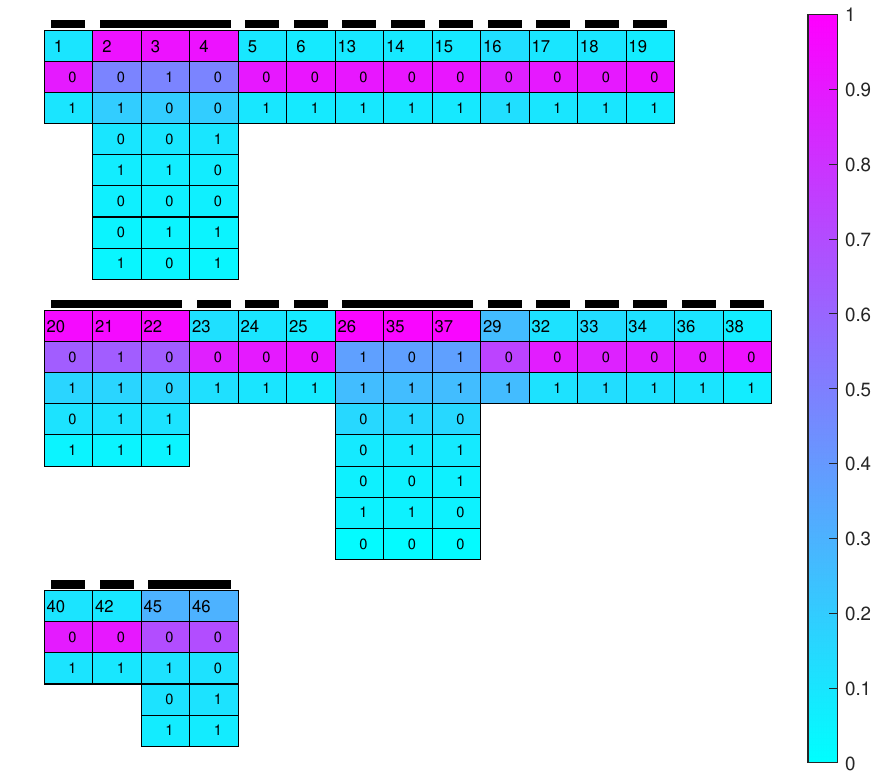} 
\end{center}
\caption{Sao Paulo traffic example: $50\%$ Cartesian credible set with $M=2$. Black lines indicate the blocks of the partition.}\label{SPT}
\end{figure}
A measure of the slowness of traffic is recorded from 7am to 8pm at half-hour intervals giving 27 measurements for each day. 
There are also 16 covariates recorded about problems in the Sao Paulo traffic network. We fit a regression model which uses a random walk to model the effect of the hour-of-day, the 16 covariates and an effect for the day of the week. 
The regression model for the  slowness of traffic of the $i$-th observation, $y_i$, is
\[
y_i = m(x_i) + \sum_{i=27}^{46} \beta_i Z_i + \epsilon_i
\]
where $x_i$ is the hour, $Z_i$ are the covariates and  day-of-the-week dummy variables and $\epsilon_i\sim \N(0, \sigma^2)$. 
We model $m(x)$ using a random walk giving regression coefficients $\beta_i = m(i + 1) - m(i)$ for $i = 1,\dots, 26$, for the $i$-th observation. 

\begin{figure}[h!]
\begin{center}
\includegraphics[scale = 0.8, trim = 50mm 120mm 60mm 120mm, clip]{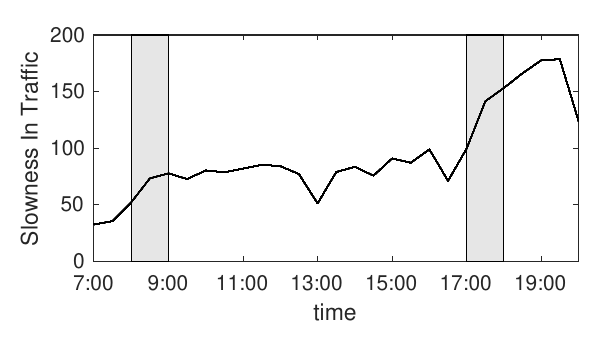} 
\end{center}
\caption{Sao Paulo traffic example: mean Slowness in Traffic as a function of time. The grey blocks indicate the times associated with regressors (2, 3, 4) and (20, 21, 22).}\label{SP1}
\end{figure}

The 50\% Cartesian credible set is shown in Figure~\ref{SPT}. There are three clear blocks of variables with blocks PIPs above 0.5 (those with variables coloured in pink). The first two blocks (2, 3, 4) and (20, 21, 22) correspond to increments in the random walk prior for $m(x)$. Figure~\ref{SP1} shows the mean slowness in traffic over time. 
 The times associated with these two blocks of regressors are indicated by grey shading and, clearly, these are the times 
with the largest increases in slowness of traffic.  The uncertainty in the variable selection for these blocks represents uncertainty about which and how many increments are needed to model these changes in the level of the response. The third block has a high posterior probability of including at least one of the variables (0.97) but there's substantial uncertainty.


\subsection{Yield of soybean}

\cite{soybean} collected data from an experiment to understand the effect of physical characteristics of cultivars of soybean on grain yields. The experiment was run in two seasons with four replicates of 40 cultivars giving a total of 320 observations. We use a linear regression model to estimates the effects of the season and the characteristics. 
\begin{table}[h!]
\begin{center}
\begin{tabular}{llll}\hline
 Variable 	& Description & Variable 	& Description\\\hline
S & Season (1 or 2) &  NGP	&		Number of legumes per plant (unit) \\
 PH	&		plant height (cm)  & 	 NGL	 &		Number of grains per plant (unit) 	\\
 IFP	&		insertion of the first pod (cm) & NS	&		Number of grains per pod (unit) \\
 NLP	&		Number of stems (unit)  & MHG &		Thousand seed weight (g) 		\\\hline
\end{tabular}
\end{center}
\caption{Soybean example: the variables in the regression}\label{t:soybean}
\end{table}
A list of the variables are given in Table~\ref{t:soybean}.
\begin{figure}[h!]
\begin{center}
\includegraphics[scale = 0.8, trim = 30mm 140mm 30mm 135mm, clip]{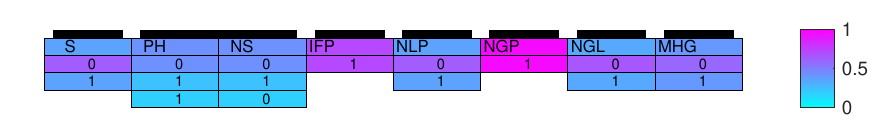} 
\end{center}
\caption{Soybean example:  $50\%$ CCS with $M=2$. Black lines indicate the blocks of the partition.}\label{f:soybean}
\end{figure}
The 50\% Cartesian credible set is shown in Figure~\ref{f:soybean}. Number of legumes per plant is clearly very important for yield (posterior probability of inclusion is 0.96) and insertion of the first pod has some evidence (posterior probability of inclusion is 0.71). For all other variables, there's no support for the inclusion of the variables.

\subsection{Student performance}

\cite{Cortez2008UsingDM} describe a study of student performance. The response is 
the final marks of students in Maths or Portuguese in two Portuguese schools. There are 30 covariates covering family background, social habits and attitudes to education. The full list of regressors are included in Appendix~\ref{a:further}.
 There are a mix of numeric (continuous), ordinal, binary and categorical covariates.
The two responses were modelled using a linear regression with  categorical covariates included using dummy variables with the first category used as a baseline. As is common  in data analysis, each regressor (including dummy variables) could be individually selected rather than allowing grouping of  dummy variables related to different levels of the same variable. 

\begin{figure}[h!]
\begin{center}
\includegraphics[scale = 0.8, trim = 30mm 105mm 30mm 105mm, clip]{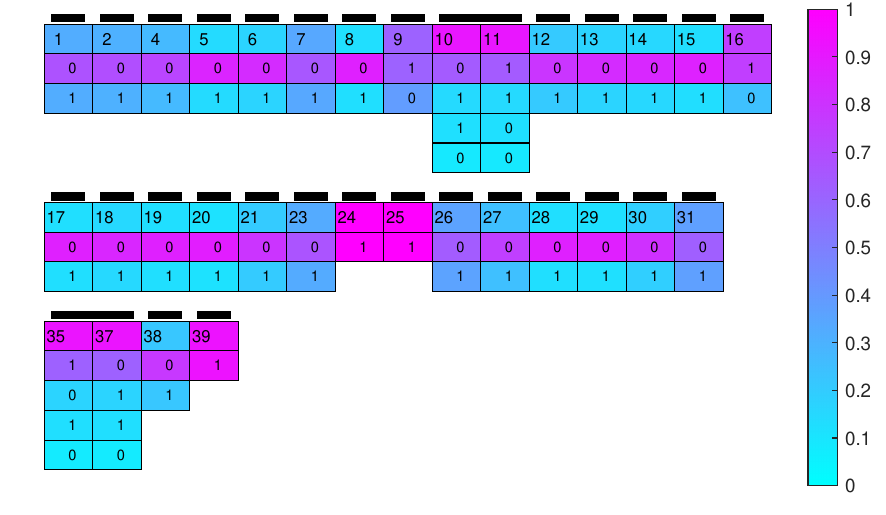} 
\end{center}
\caption{Student performance example:  $50\%$ Cartesian credible set with $M=2$ when final mark in Maths is used as the response.
Black lines indicate the blocks of the partition.}\label{f:d1_sub}
\end{figure}
The 50\% Cartesian credible set when the response is the final marks in Maths is shown in Figure~\ref{f:d1_sub}.  Three variables have strong support as important: the number of failures in past classes (24, PIP = 1.00), whether the student receives extra educational support (25, PIP = 1.00) and the number of absences (39, PIP = 0.93). It seems reasonable that these variables should have an effect on  educational performance. There is  also some support for the inclusion of 
``Father's job is other'' (16, PIP = 0.75) and mother's job   is healthcare (9, PIP = 0.62). Two blocks of variables are also identified. Firstly, the difference between baseline (teacher) and several levels of mother's job  (civil service, or  ``at home'') (10, 11). The block PIP  is 0.91. The highest probability submodel only includes ``at home'' (11, marginal posterior probability is 0.64).
The second block is how often a student goes out with friends and their weekend alcohol consumption (35, 37). The posterior probability of including both is 0.92.  It's not surprising that these two variables are hard to disentangle since socialising can involve alcohol consumption. The highest probability submodel only includes going out (35, marginal posterior probability is 0.62).

\begin{figure}[h!]
\begin{center}
\includegraphics[scale = 0.8, trim = 30mm 90mm 30mm 90mm, clip]{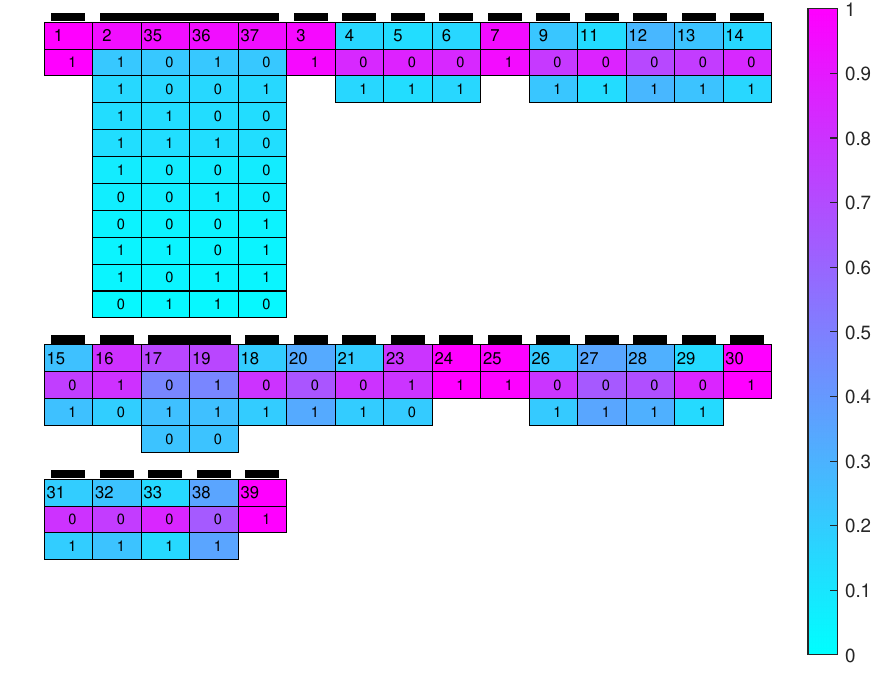} 
\end{center}
\caption{Student performance example:  $50\%$ Cartesian credible set with $M=2$ when final mark in Portuguese is used as the response.
 Black lines indicate the blocks of the partition.}\label{f:d2_sub}
\end{figure}
The  50\% Cartesian credible set when the response is the final marks in Portuguese is shown in Figure~\ref{f:d2_sub}. Some variables which are important for Maths are also important for Portuguese:
the number of failures in past classes (24, PIP: 1.00), whether the student receives extra educational support (25, PIP: 0.99), the number of absences (39, PIP: 1.00), and father's job is ``other'' (16, PIP: 0.81). The block related social activities in Maths results expands  to include sex (2) and the daily alcohol consumption (36) (the block PIP is 0.95). There's a lot of uncertainty over the sub-model with largest  posterior probability for a single sub-model being 0.22.
In addition, the school (1, PIP: 1.00), age (3, PIP: 0.96), mother's education (7, PIP: 0.95), whether the students wishes to take higher education (30, PIP: 1.00), and study time (23, PIP: 0.86). There's also a block containing the reason for choosing the school (reputation, 17, or other, 19) which posterior probability of 0.72 that one of the variables is included.

\subsection{GWAS example: Systemic Lupus Erythematosus}

We consider a GWAS for Systemic Lupus Erythematosus using a case-control study. The 
 presence or absence of the disease is the binary outcome in a logistic regression with the SNPs as explanatory variables.
 It is assumed that there are a number of causal SNPs which truly cause biological differences between the case and controls. A simple analysis would fit a logistic regression separately for each SNP  and declare a SNP as associated with the outcome if it has a p-value smaller than a chosen (typically, extremely small) threshold.  However, due to correlation in the genome, we usually see many SNPs declared as associated for each causal SNP. We refer to this group of SNPs as the signal. The goal of the analysis is to determine the number of independent signals and which SNPs are most likely the causal SNPs at each signal. 

\begin{figure}[h!]
\begin{center}
\includegraphics[trim = 40mm 110mm 40mm 110mm, scale = 1.0, clip]{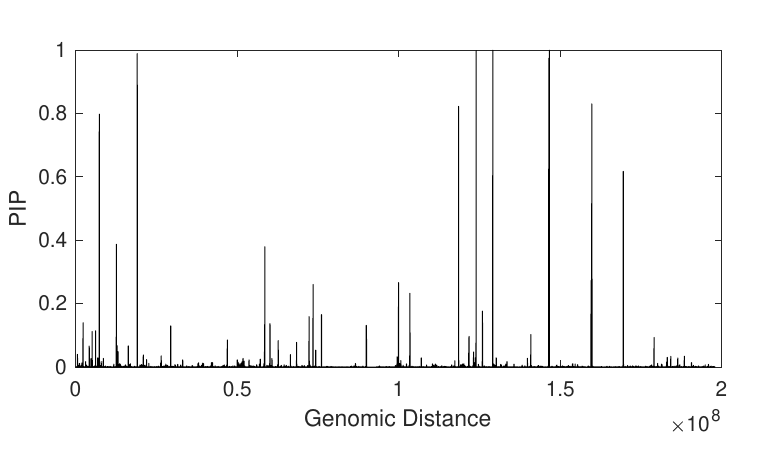}
\end{center}
\caption{GWAS example: Marginal PIPs for chromosome 3.}\label{f:GWAS_PIP}
\end{figure}

 The data  contains 4,036 cases and 6,959 controls, with the cases and controls having European ancestry. 
In this example we only look at SNPs on Chromosome 3 ($42, 430$ SNPs/variables). We apply Bayesian variable selection to the data using a logistic regression model with all SNPs as potential covariates and fit the model using an adaptive MCMC algorithm \citep{griwan21}. The full specification of the Bayesian model is provided in Appendix~\ref{appendix:setup}.
Figure~\ref{f:GWAS_PIP} shows the PIPs for each SNP. There are  several clear signals from SNPs with PIPs equal to 1 and others with large PIPs.
The 50\% Cartesian credible set is found by first retaining only those SNPs with a PIP above 0.04 (leaving 69 variables). The genomic distances of these 69 variables are given in  Appendix~\ref{a:further}.  
The median model contains 13 variables which using the indices of the retained SNPs are 
5, 10, 37, 45, 47, 48, 50, 51, 52, 53, 54, 58, 63.


\begin{figure}[h!]
\begin{center}
\includegraphics[trim = 20mm 90mm 15mm 80mm, scale = 0.9, clip]{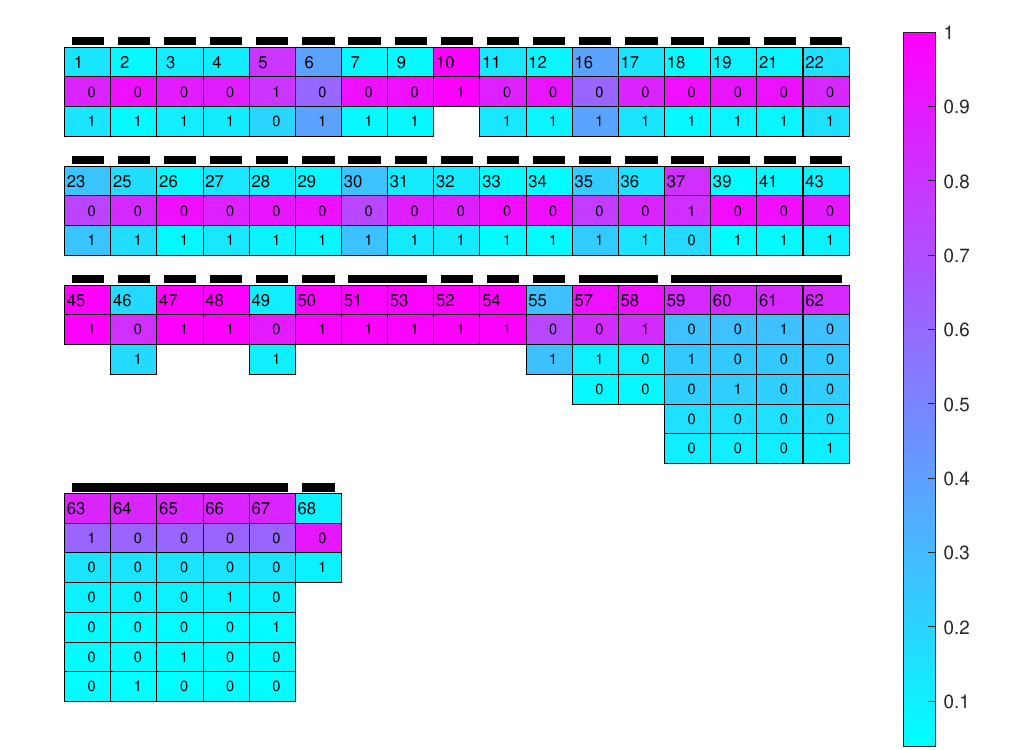}\
\end{center}
\caption{GWAS example: $50\%$ Cartesian credible set with $M=2$.  Black lines indicate the blocks of the partition.}\label{f:GWAS_2}
\end{figure}
The 50\% Cartesian credible set is shown in Figures~\ref{f:GWAS_2}.. The most striking aspect is that the blocks of variables are contiguous which reflects the  serial correlation of the SNPs.
The results  show that 11 variables in the median model do not include the possibility of exclusion in the credible set 
(10,  45, 47, 48, 50, 51, 52, 53, 54) and two variables are included as singletons with the possibility of exclusion but stronger support for inclusion (5 and 37). 
Two other variables from the median model are included as the most important submodels in blocks which have strong support for inclusion. These are 58 with 57 and 63 with 64, 65, 66, and 67. 
In addition, there is one block with strong support (and, which has no variables included in the median model) containing 59, 60, 61, and 62 (genomic distance: 159767453--159804903), 
which has block PIP equal to 0.84 (the posterior probability of modal sub-model for this block is 0.28).


\section{Discussion}

Cartesian credible sets are a useful way to summarize and visualize the output from Bayesian variable selection. The sets are constructed by finding a partition of the variable and an approximation to the  posterior distribution where variables in different subsets are independent. Credible sets are defined as a Cartesian products of sets of sub-model and can be found using a  simple heuristic algorithm which 
minimizes the KL divergence between the posterior and the approximation and chooses  a parameter controlling the level of approximation and this can be chosen using a criterion which trades-off the accuracy of the approximation and overall the size of the credible set. 
The method can be run quickly as post-processing of an MCMC sample from the posterior distribution on the models. The results can be visualized by plotting the list of sub-models for each subset and coloured according to the posterior probability of the sub-models and the posterior probability of including at least one variable in a subset. This gives a much more complete picture of the posterior distribution than the usually plotted PIPs and helps understanding of the interchangeability of variables which are highly collinear.
This allows us to avoid dilution effects from posterior mass being distributed across such variables by considering blocks of variables. 

The approach suggests several different directions for future research in Bayesian variable selection. The approximation of the posterior distribution from Bayesian variable selection using independent blocks of variables can be useful in computational methods. Some recently developed algorithms for high-dimensional variable selection use an independent approximation to the posterior distribution. In MCMC methods, the ASI algorithm \citep{GriLatSte21}, and its development to ARNI and PARNI  \citep{LiaLivGri22} use an independent approximation. Similarly, \cite{RaySza22} study variational inference for Bayesian variable selection where the variational distribution is independent across the variables. In both cases, the methods developed in this paper can act as starting point for defining iterative algorithms which learn the main features of the correlation structure in the posterior distribution. In this paper, we argue that a Cartesian credible set is a more useful summary of the posterior distribution than the smallest credible set. However, if the smallest credible set is interesting and the probability of individual models can be well-approximated by MCMC output, a Cartesian credible set could be used as a starting point. The smallest credible set can then be found by iteratively removing model with the smallest posterior probability. 

Summarizing Bayesian variable selection is challenging as the posterior distribution is defined on a discrete space. Defining the credible set as  a Cartesian product of sub-models on blocks of variables combines with an independent approximation of the posterior distribution to provide a simple summary method. There are many other Bayesian methods where the variable of interest is high-dimensional and discrete. For example, in Bayesian nonparametrics, clustering using mixture models and feature allocation or change-point models in time series. If an independent approximation is appropriate in these models, then the methods developed in this paper could be easily applied. In posterior distributions with more structure, other representations such as factor models could be used to define credible sets in these more complicated settings.

\section*{Acknowledgements}

I would like to thank Xitong Liang and Dr Sam Livingstone from University College London and Professor Mike Smith from Melbourne Business School for helpful discussions.
I would also like to thank Professor Tim Vyse and Dr David Morris from King's College London for providing the data in the GWAS example and initial motivation to consider effectively expressing model uncertainty.

 \bibliographystyle{chicago}
\bibliography{ref}

\appendix

\section{Proofs}

\subsection{Proposition 1}

\begin{align*}
p(S\vert \mbox{Data}) 
&= \sum_{\gamma \in S} p(\gamma\vert \mbox{Data})
=  \sum_{\gamma \in \otimes_{i=1}^L S_i} \prod_{i=1}^L p(\kappa_i\vert \mbox{Data})
=  \sum_{\kappa_1 \in  S_1} \dots  \sum_{\kappa_L \in  S_L}  \prod_{i=1}^L p(\kappa_i\vert \mbox{Data})\\
&=  \prod_{i=1}^L \sum_{\kappa_i \in  S_i}    p(\kappa_i\vert \mbox{Data})
=  \prod_{i=1}^L   p(S_i\vert \mbox{Data})
\end{align*}

\section{Bayesian inference in regression models}\label{appendix:setup}

\subsection{Linear regression}

We fit a linear regression model where $\alpha$ represents the intercept, $\beta_{\gamma}$ represents the regression coefficients for model $\gamma$, and $\sigma^2$ represents the observation variance.  The prior distributions are
\[
 p(\alpha, \sigma^2) \propto \sigma^{-2},  \quad \beta_{\gamma} \sim\N(0, g\,I_{p+1}), \quad
\gamma_i \stackrel{i.i.d.}{\sim} \mbox{Bernoulli}(\pi), \ i = 1,\dots, p, 
\]
\[
\pi\sim\mbox{Beta}\left(1, \frac{p - p_0}{p_0}\right), \quad g\sim\mbox{Half-Cauchy}.
\]
The posterior distribution was sampled using a simple Add-Delete-Swap sampler for $\gamma$ with Gibbs steps to update $\pi$ and $g$ and all other parameters integrated out. The samplers were run to collect 100 000 iterations
 with a burn-in of  100 000 and every  300th value recorded (total number of iterations 30 100 000).

\subsection{Logistic regression}

We fit a logistic regression model where $\alpha$ represents the intercept, and $\beta_{\gamma}$ represents the regression coefficients for model $\gamma$.  The prior distributions are
\[
  \beta_{\gamma} \sim\N(0, I_{p+1}), \quad
\gamma_i \stackrel{i.i.d.}{\sim} \mbox{Bernoulli}(\pi), \ i = 1,\dots, p, \quad
\pi\sim\mbox{Beta}\left(1, \frac{p - p_0}{p_0}\right),
\]
The posterior was sampled using the ASI sampler for logistic regression from 
\cite{griwan21}. Three independent runs were used with each run containing 25 replicate chains. Each chain was run for 15 000 iterations with a burn-in of 5 000 iterations and the subsequent 10 000 samples thinned every 10th value. This gave 1 000 samples for each replicate chain, 25 000 samples for each run, and  75 000  samples overall.

\section{\cite{GeoMcC97} example}\label{a:GM}

This example has $n = 180$ and $p = 15$. Let $Z\sim\N(0, I_p)$ and $Z_0\sim \N(0, 1)$ and the variables 
are  $X_1 = Z_1 + 2  Z_0$, $ X_3 = Z_3 + 2  Z_0$, $X_5 = Z_5 + 2  Z_0$,
$X_8 = Z_8 + 2  Z_0$, $X_9 = Z_9 + 2  Z_0$,
$X_{10} = Z_{10} + 2  Z_0$, $X_{12} = Z_{12} + 2  Z_0$, 
$X_{13} = Z_{13} + 2  Z_0$, $X_{14} = Z_{14} + 2  Z_0$
$X_{15} = Z_{15} + 2  Z_0$,  $X_2 = X_1 + 0.15  Z_2$,
$X_4 = X_3 + 0.15  Z_4$,
$X_6 = X_5 + 0.15  Z_6$, $X_7 = X_8 + X_9 - X_{10} + 0.15  Z_7$,  and
$X_{11} = X_{14} + X_{15} - X_{12} - X_{13} + 0.15  Z_{11}$.
The regression coefficients are 
$\beta = (1.5, 0, 1.5, 0, 1.5, 0, 1.5, -1.5, 0, 0, 1.5, 1.5, 1.5, 0, 0)'$ and so the non-zero regression coefficients are
1, 3, 5, 7, 8, 11, 12, 13.

\section{Further details of applications}\label{a:further}

The variable in the ``behaviour of traffic in Sao Paulo, Brazil'' example are shown in Table~\ref{t:SP_vars}
\begin{table}[h!]
\begin{center}
\begin{tabular}{llllll}\hline
No. & Variable & No. & Variable & No. & Variable\\\hline
27 &	Immobilized Bus & 34 &	 Incident Involving  &41 &	Semaphore Off\\
& & & Dangerous Freight & \\
28 &	Broken Truck &35 &	Lack Of Electricity &42 &	Intermittent Semaphore \\
29 &	Vehicle Excess &36 &	Fire &43 &	15th December \\
30 &	Accident Victim &37 &	Point Of Flooding &44 &  16th December \\
31 &	Running Over &38 &	Manifestations &45 & 17th December \\
32 &	Fire Vehicles   &39 &	Defect In The Network  &46 & 18th December\\
& & & of Trolleybuses & & \\
33 &	 Occurrence Involving Freight & 40 &	Tree On The Road & &\\\hline
\end{tabular}
\end{center}
\caption{The regressors for the Sao Paulo traffic example}\label{t:SP_vars}
\end{table}
In the GWAS example, the variable used after PIP screening are shown in Table~\ref{t:GWAS_picked}
\begin{table}[h!]
{\small
\begin{center}
\begin{tabular}{cccccccccccc}\hline
No. & Distance & No. & Distance & No. &Distance& No. & Distance & No. & Distance & No. & Distance \\\hline
1 & 2233338 &  13  & 58377159 &   25 & 76049395 &  37 & 118468553 & 49 & 140867938 &  61 & 159791628\\
 2 &   4137393 &   14 & 58382846 &  26 & 76077143 &  38 & 121618371 & 50 & 146419285 & 62 & 159804903\\
 3  &  5038795 &  15 &  58405947 &  27  & 89904882 & 39 & 121620787 & 51 & 146579888 & 63 & 169461571\\
 4  &  6104625 & 16 &  58512237 &  28 & 89982853 & 40 & 121632432 & 52 & 146594746 & 64 & 169492101\\
 5 &   7255106 &  17  & 60095604 & 29  & 99905803 &  41& 121646886 &  53 &  146601081 &  65 & 169497585\\
 6 &  12531065 & 18 &  60098968 & 30 &  99906993 &42 & 121664112 &  54 & 146601295 & 66 & 169508272\\
 7 &  12746807 & 19 &  62645954 & 31 &  99914139 &   43 & 121715319 &  55 & 159501673 &  67 & 169512145\\
 8  & 13052644 &  20 &  68324960 & 32 & 103387440 &  44 & 123131254 &56  & 159582382 & 68 & 178995657\\
  9  & 16224346 &  21 & 68328854 & 33 & 103388364 & 45 & 123925271 &  57 & 159728987 &  69 & 178997778\\
 10 &  18998569 & 22 &  72187269 & 34 & 103394499 &  46 & 125880208 & 58 & 159732983 & & \\
  11 &  29378090 & 23  & 73444439 & 35 & 103404111 & 47 & 129083281 &  59 & 159767453 & & \\
 12  &  46880128 & 24 &  74270459 & 36 & 103433922 & 48 & 129084581 & 60 & 159780373 & & \\\hline
\end{tabular}
\end{center}}
\caption{GWAS example: the genomic distance of each variable chosen using the PIP screening}\label{t:GWAS_picked}
\end{table}
The variable in the ``student performance'' example are shown in Table~\ref{t:Student_vars}
\begin{table}[h!]
\begin{tabular}{rll}\hline
Number & Variable & Type \\\hline
1 & school - student's school (Gabriel Pereira or Mousinho da Silveira)  & Binary\\ 
2 & sex - student's sex (female or male) & Binary \\
3 & age - student's age   & Numeric\\
4 & address - student's home address type (urban or rural)  & Binary\\
5 & famsize - whether family size is  less or equal to 3  & Binary\\
6 & Pstatus - parent's cohabitation status (living together or  apart) & Binary\\
7 & Medu - mother's education  & Ordinal\\
 &  (0 - none,  1 - primary education (4th grade), 2 – 5th to 9th grade, & \\
 & 3 – secondary education or 4 – higher education) & \\
8 & Fedu - father's education & Ordinal \\
 &  (0 - none,  1 - primary education (4th grade), 2 – 5th to 9th grade, & \\
 & 3 – secondary education or 4 – higher education) & \\
9--12 & Mjob - mother's job & Categorical\\
& (teacher, healthcare related, civil services, at home, other) & \\ 
13--16 & Fjob - father's job  & Categorical \\
& (teacher, healthcare related, civil services, at home, other) & \\ 
17--19 & reason - reason to choose this school  & Categorical \\
& (close to home, school reputation, course preference, other) \\
20--21 & guardian - student's guardian (mother, father, other) & Categorical \\
22 & traveltime - home to school travel time  & Ordinal\\
& (1 - $<15$ min., 2 - $15$ to $30$ min., 3 - $30$ min. to $1$ hour, or 4 -$ >1$ hour) \\
23 & studytime - weekly study time & Ordinal \\
& (1 - $<2$ hours, 2 - $2$ to $5$ hours, 3 - $5$ to $10$ hours, or 4 - $>10$ hours) \\
24 & failures - number of past class failures (capped at 4) & Numeric\\
25 & schoolsup - extra educational support (yes/no) & Binary \\
26 & famsup - family educational support (yes/no) & Binary \\
27 & paid - extra paid classes within the course subject  & Binary\\
& (Math or Portuguese) (yes/no)  & \\
28 & activities - extra-curricular activities (yes/no) & Binary \\
29 & nursery - attended nursery school (yes/no) & Binary \\
30 & higher - wants to take higher education (yes/no) & Binary \\
31 & internet - Internet access at home (yes/no) & Binary \\
32 & romantic - with a romantic relationship (yes/no) & Binary \\
33 & famrel - quality of family relationships (1 - very bad to 5 - excellent) & Ordinal \\
34 & freetime - free time after school (1 - very low to 5 - very high) & Ordinal \\
35 & goout - going out with friends (1 - very low to 5 - very high) & Ordinal  \\
36 & Dalc - workday alcohol consumption (1 - very low to 5 - very high) & Ordinal \\
37 & Walc - weekend alcohol consumption (from 1 - very low to 5 - very high) & Ordinal \\
38 & health - current health status (from 1 - very bad to 5 - very good) & Ordinal \\
39 & absences - number of school absences & Numeric \\\hline
\end{tabular}
\caption{Variables used in the student performance example}\label{t:Student_vars}
\end{table}

\end{document}